\documentclass[a4paper]{article}
 \usepackage{psfrag}
\usepackage{graphicx}
\usepackage{amsmath}
\usepackage{hyperref}
\usepackage{color}

\begin{document}

\begin{center}

{\bf \Large Situations in traffic - how quickly they change}\\[5mm]

{\large M. J. Krawczyk$^{1a}$, C. Beltran Ruiz$^{2b}$ and K. Ku{\l}akowski$^{1c}$ }\\[3mm]

{\em

$^1$Faculty of Physics and Applied Computer Science, AGH - UST, al. Mickiewicza 30, PL-30059 Krak\'ow, Poland\\
$^2$Sociedad Iberica de Construcciones Electricas, C/Sep\'ulveda 6, 28108 Alcobendas, Madrid, Spain

}

{\tt $^a$gos@fatcat.ftj.agh.edu.pl, $^b$cbeltran@sice.com, $^c$kulakowski@novell.ftj.agh.edu.pl}

\today

\end{center}

\begin{abstract}
Spatio-temporal correlations of intensity of traffic are analysed for one week data collected in the motorway M-30 around Madrid in January 2009.
We found that the lifetime of these correlations is the shortest in the evening, between 6 and 8 p.m.  This lifetime is a new indicator how much 
attention of drivers is demanded in given traffic conditions.
\end{abstract}

\noindent

{\em PACS numbers:} 

\noindent

{\em Keywords:} traffic, data analysis, lifetime

\section{Introduction}

As an interdisciplinary application of statistical physics, physics of traffic is not new. As several problems posed there remain to be solved, 
the topic is far from being exhausted. One of central questions is how traffic jams arise, with another question - how to control the jams - in the context. Around these issues, a lot of modelling approaches have been proposed \cite{chow,hel,kai,mae}, from partial differential equations to cellular automata. As a rule, two phases of traffic are discussed: free flow and jam. Apart from these two, a synchronized phase has been distinguished \cite{kerre,ker}. Later, the concept was criticized as ambiguous \cite{hell}. According to this critique the effect when vehicles move "nearly synchronized in different lanes of the highway" \cite{kerre} appears in qualitatively different states and therefore it cannot be treated as a criterion of a specific phase. Still, the synchronization as an effect (not a phase) has been frequently called in \cite{h1998,lee,lub,h2004}. In particular, two synchronized modes of traffic have been distinguished in \cite{lub}: lightly and heavily congested traffic. \\

Our concern here is devoted to the lifetime of spatial configurations of moving vehicles. The motivation is, most briefly, that these lifetimes give information on the rates of transitions of the traffic system from one state to another. Specific configurations of vehicles with their velocities can be treated as microscopic states. This kind of data is hardly available; more often coarse-grained data are collected as intensities $A$ averaged over some time periods, 30 seconds at best. From readings $A_1(t)$ and $A_2(t)$ collected with two sensors $1,2$ of distance $x$ one from another, we evaluate the time average of the correlation of $A_1(t)$ with $A_2(t+\tau)$. Once the coarse-grained state changes only slightly in time, the correlation should show a maximum at the delay $\tau=x/v$, where $v$ is the mean velocity of vehicles. Further, the velocity spread of vehicles is expected to lead to a finite length of spatial correlations of $A$. The latter rescaled again by $v$ is a measure of lifetime of the coarse-grained states. An obvious shortcoming of this procedure is that the time of measurement (1 minute) is not commensurate with the mean time $x/v$ of motion of vehicles from one sensor to another. However, this difficulty can be evaded at least partially by averaging over more pairs of sensors.\\

The lifetime seems to be a new tool, and in particular we are not aware about any research on the spatio-temporal correlations of traffic flow. In \cite{neu}, the autocorrelation was investigated of the local density, the average velocity and the flow for the free-flow, synchronized and congested phases. The autocorrelation functions measured with one sensor showed a slow decrease of density and flow in the free flow state, and much quicker decrease in the synchronized state, while in the congested (stop-and-go) state, the autocorrelation functions oscillate in time (Figs. 10 and 12 in \cite{neu}). According to \cite{ker}, the sequence of phases from free flow to jammed phase is through the synchronized phase. On the other hand, an exemplary time series with a transition from the free flow to the congested phase, shown in \cite{neu} (Fig. 13), show a sharp character of the transition, with only momentary increase of the variance of velocity. Both the short time of the transition and the short timescale of the autocorrelation functions can be due to the metastable character of the synchronized state. The data on the lifetime should provide information how quickly the traffic states change. \\

In the next section we give the details on our traffic data. Two subsequent sections are devoted to the details of the calculation and to the numerical results. Last section contains our conclusions.\\

\begin{figure}[ht]
 \centering {\centering \resizebox*{12cm}{9cm}{\rotatebox{-90}{\includegraphics{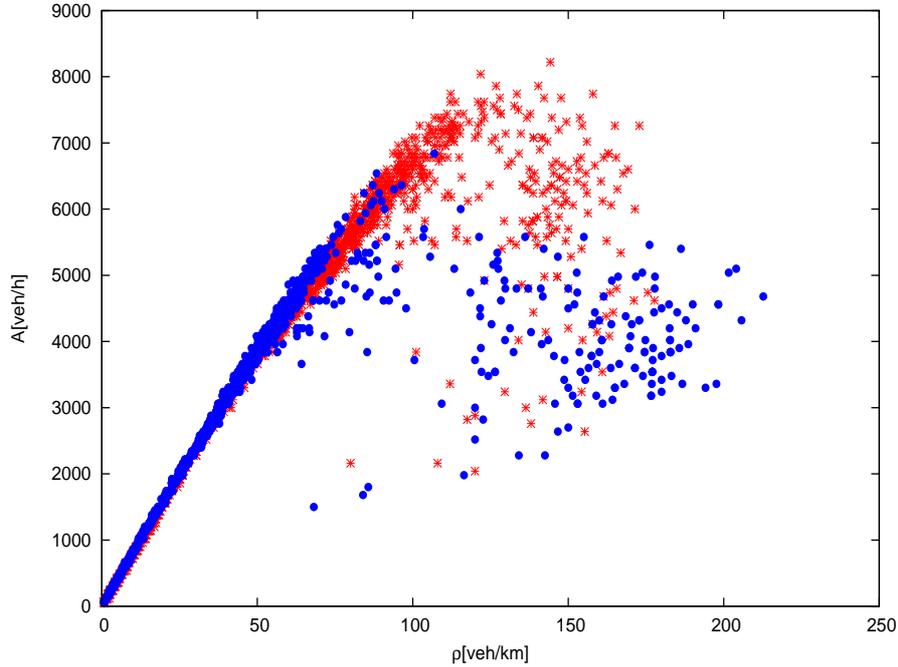}}}}
\caption{The fundamental diagram for two sensors (marked as $\ast$ and $\bullet$) in the motorway M-30.}
\label{fig-3}
\end{figure}

 \begin{figure}[ht]
 \centering {\centering \resizebox*{12cm}{9cm}{\rotatebox{-90}{\includegraphics{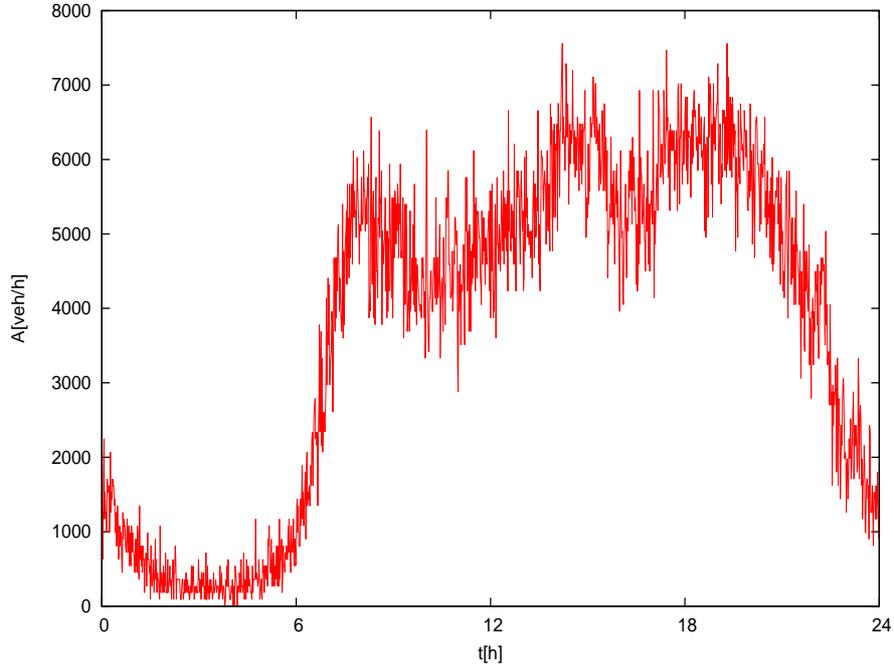}}}}
\caption{Intensity against time as read with one sensor by 24 h in the motorway M-30 (exemplary curve).}
\label{fig-4}
\end{figure}

 \begin{figure}[ht]
 \centering {\centering \resizebox*{12cm}{9cm}{\rotatebox{-90}{\includegraphics{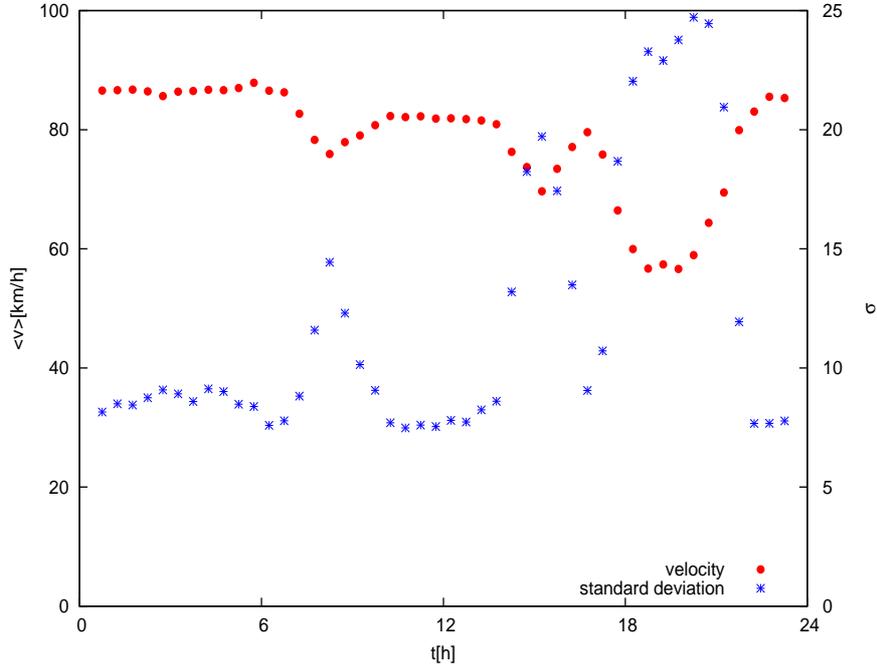}}}}
\caption{Velocity (red dots, left scale) and its standard deviation (blue stars, right scale) against time as read with one sensor by 24 h in the motorway M-30.}
\label{fig-5}
\end{figure}

 \begin{figure}[ht]
 \centering {\centering \resizebox*{12cm}{9cm}{\rotatebox{-90}{\includegraphics{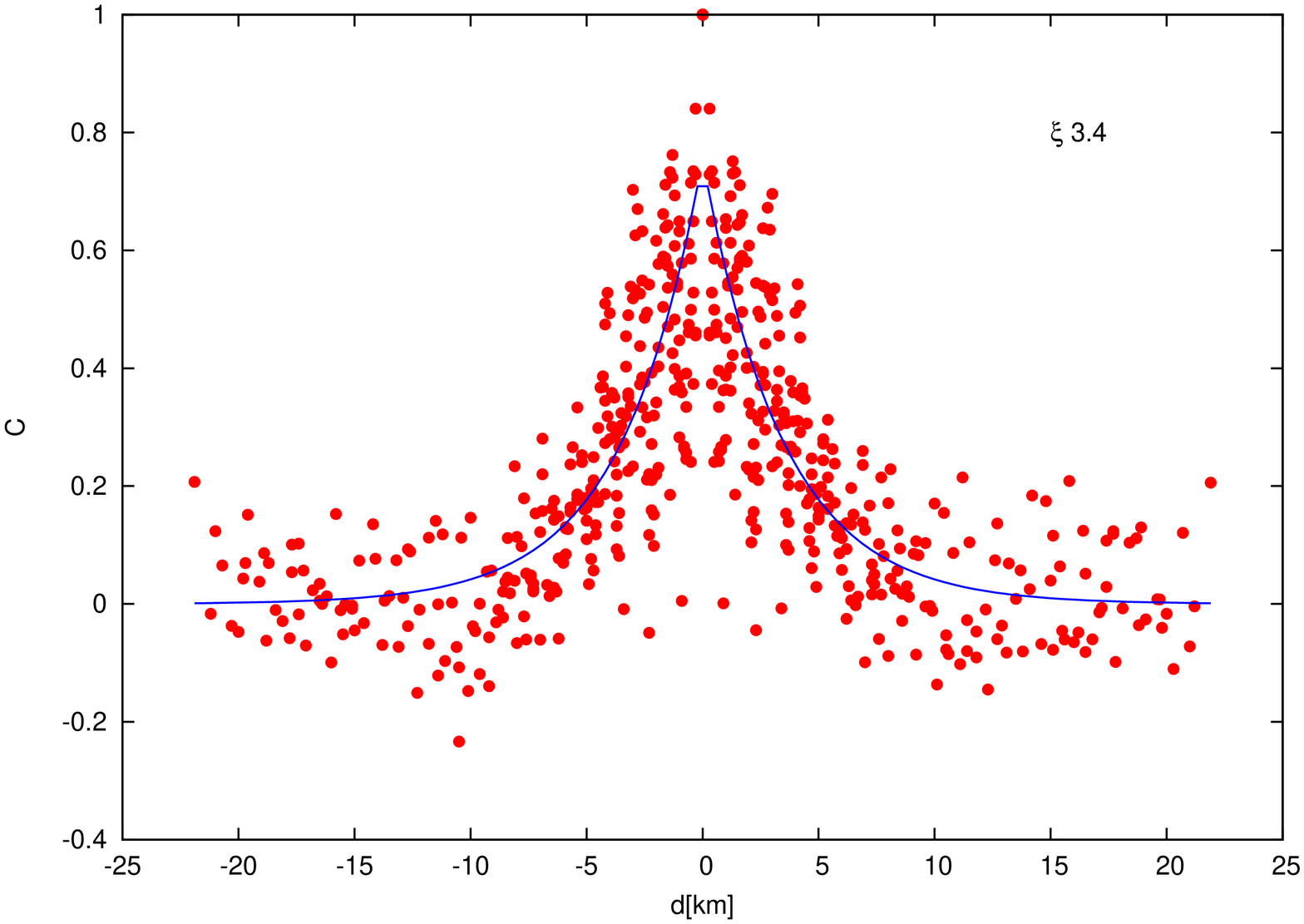}}}}
\caption{Correlation against distance - an average over workdays for all pairs of 24 sensors between 4 and 4.30 a.m. in the motorway M-30.}
\label{fig-6}
\end{figure}

 \begin{figure}[ht]
 \centering {\centering \resizebox*{12cm}{9cm}{\rotatebox{-90}{\includegraphics{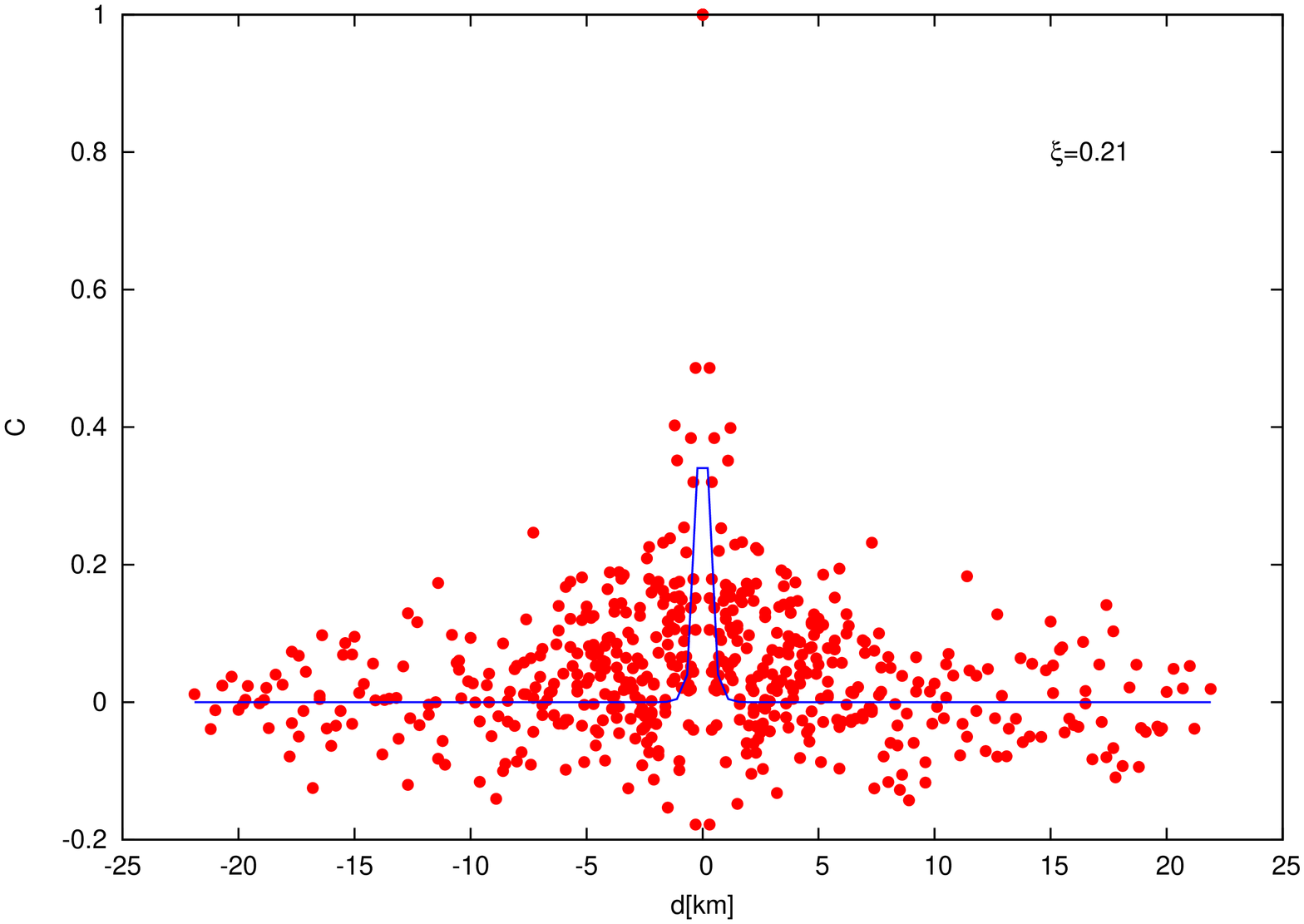}}}}
\caption{Correlation against distance - an average over workdays for all pairs of 24 sensors between 6 and 6.30 p.m. in the motorway M-30.}
\label{fig-7}
\end{figure}

 \begin{figure}[ht]
 \centering {\centering \resizebox*{12cm}{9cm}{\rotatebox{-90}{\includegraphics{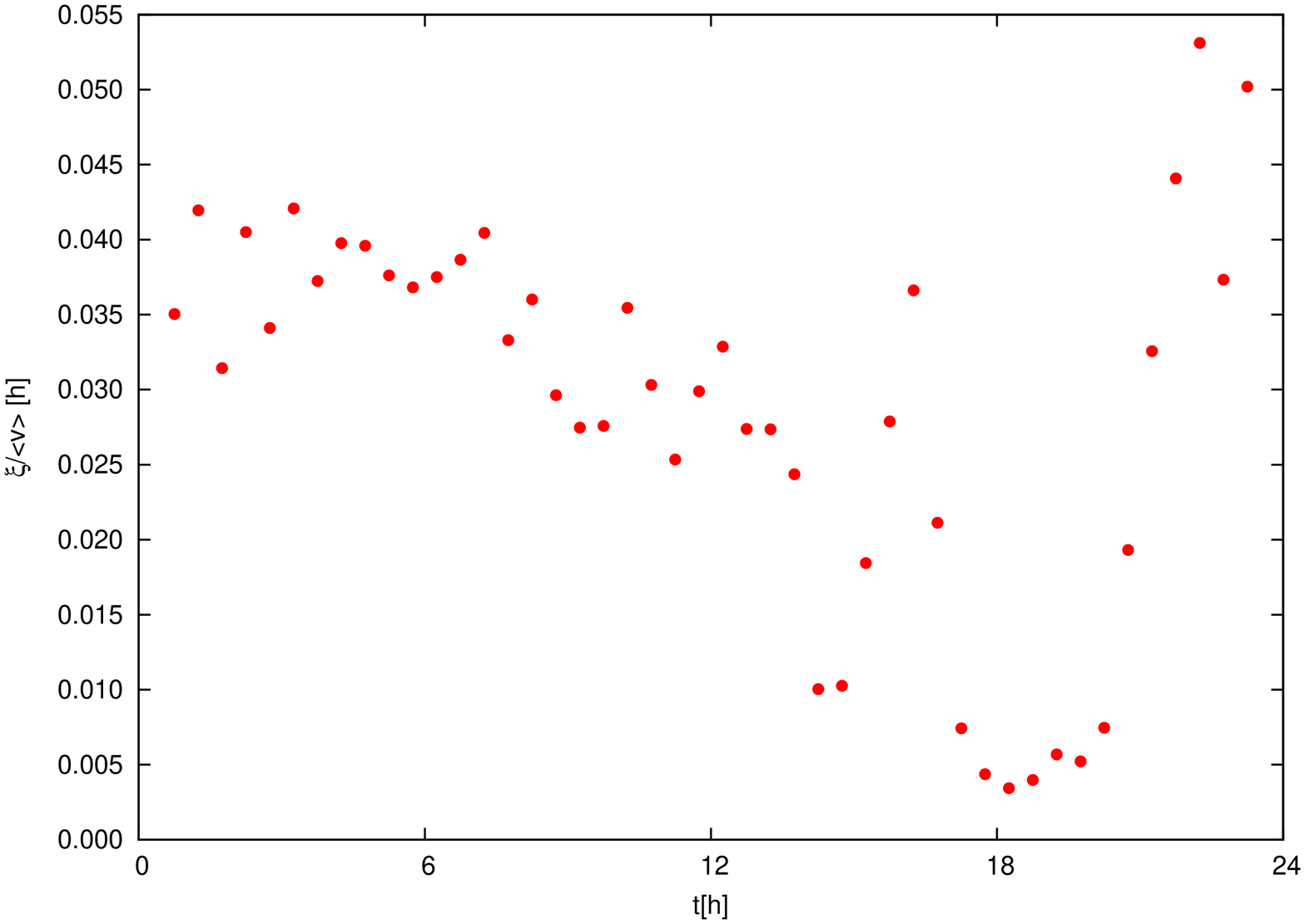}}}}
\caption{Lifetime $\tau=\xi/<v>$ against time.}
\label{fig-8}
\end{figure}

\section{Data}

The data are collected from 1-minute aggregated readings of 24 sets of sensors –or measurement points- placed along a 13 km-length road stretch of the M30 motorway that surrounds the center of Madrid. Data sets cover a period of one week readings, between January 26th and February 1st, 2009.
These real traffic datasets provided by the detectors include (i) counting of vehicles; (ii) intensity or flow of traffic (vehicles/h); (iii) traffic density (vehicles/km); (iv) occupancy data (time that the sensor is detecting the presence of a vehicle); (v) real time vehicle speed (km/h); and (vi) vehicle length and classification by categories (car, bus, van, truck, etc.). A detailed description of the data can be found in \cite{linz}.\\

Real world traffic data sets from M30 motorway have been measured by a detection system based on inductive loops; these inductive sensors are used in detection systems that rely on the principle that a moving magnet –a vehicle- will induce an electrical current in a nearby conducting wire, fact that is applied for the detection of vehicle presence indicators and thus for getting information of real traffic data.
The base of these vehicle detection systems is the detection of the induction change when a metallic mass –a vehicle- passes through a loop installed under the road pavement. The loop forms part of circuits that oscillate to a determined frequency that is changed by the variations of the loop´s inductance as the metallic mass passes through it. 
Then the detector will translate that physical reality (the presence of a vehicle) into an electrical signal that will be processed afterwards by a microprocessor that will digitally treat the obtained inputs transmitted by the induction loops so as to obtain real traffic data.
Detectors can aggregate real data upon different time basis. In the case of the M30 motorway detection system, traffic data are aggregated on 1-minute integration period basis.\\

Inductive loops-based detectors are considered as most accurate electronic devices for vehicle detection; they are normally used in traffic detection due to their high robustness and reliability, their low sensitivity to adverse meteorological conditions and their low cost. Therefore, the probability of erroneous vehicle sensing is very low, although they can also suffer from shifts in their reference frequency, mainly due to big thermal changes in the pavement, and from possible breaks or wire short-circuits.

\section{Calculations}

At the beginning, the fundamental diagram of the data is constructed. This diagram express the intensity of traffic as a function of vehicles density. The density is calculated from the data as a quotient of intensity $A$ and velocity $v$ at each time point.\\

For the remaining analysis, the curve showing the dependence of the number of vehicles as a function of time is subjected to detrendisation \cite{detr}. This procedure is performed as follows: For each sensor and day the whole data set is divided into one hour long subsets. Each subset is fitted by a cubic function, using the ''least square method.'' Subsequently, from each measured value, a value calculated on the basis of parameters obtained from the fitting procedure is subtracted. The procedure is applied if the amount of data in a given time period is higher than the number of parameters of the function used for fitting. Data prepared in this way is used in further analysis.

We estimate the travel time between each pair of sensors ($s_i, s_j$), on the basis of the distance between them and the average velocity of the vehicles on the given portion of the road. After the estimated time, the same vehicles which are detected near the sensor $s_i$ should appear near the sensor $s_j$. Taking into account this time shift $\tau$, the Pearson correlation coefficient $c$, between a given pair of sensors, is calculated:
\[
c_{ij}=\dfrac{\sum\limits_{t=1}^n(x_{it}-\bar{x}_i)(x_{j(t+\tau)}-\bar{x}_j)}{\sqrt{\sum\limits_{t=1}^n\left(x_{it}-\bar{x}_i\right)^2}\sqrt{\sum\limits_{t=1}^n\left(x_{j(t+\tau)}-\bar{x}_j\right)^2}}
\]
where: $i$, $j$ - items, $n$ - length of the time series, $\bar{x}_i=\dfrac{1}{n}\sum\limits_{t=1}^nx_{it}$ $\left(\bar{x}_j=\dfrac{1}{n}\sum\limits_{t=1}^nx_{j(t+\tau)}\right)$ - average value. Obtained values $c_{ij}$ lie in the range $[-1,1]$, where $-1$ - means full anti-correlation, $0$ - lack of correlation, and $1$ - full correlation. Each value is averaged with all other values obtained in the following half-hour.\\

As it is visible in the data, and what is intuitively understandable, observed traffic data is different on workdays and weekends. Because of that, in further calculation only data from workdays was taken into account. For all pairs of sensors and time periods, obtained values of the correlation coefficients are averaged over workdays. Collecting all obtained data for half-hour periods of time, we obtain a curve which shows how the correlation coefficient depends on the distance between sensors. Fitting this curve by a function $f(x)=\lambda\exp(-x/\xi)$ we obtain a correlation length $\xi$ for each period of time.

\section{Results}

In Fig.\ref{fig-3} we show the fundamental diagram of the collected data. The overall character of the obtained plot is similar to the fundamental diagrams reported in literature \cite{chow,hel,kai}. The flow intensity against time, is shown in Fig.\ref{fig-4}. The data show that the character of the curve is the same for all working days, see also the plots in \cite{linz}. In Fig.\ref{fig-5} we show the velocity and its standard deviation, averaged over 30 minutes. As we see, low velocity is correlated with its large variance.  \\

In Figs. \ref{fig-6} and \ref{fig-7}, the plots show the data on the correlation coefficient $C$. These results collected for all pairs of sensors are averaged over workdays and fitted to see the distance dependence of  $f(d)=\lambda\exp(-d/\xi)$. The value of the obtained correlation length $\xi$ is 3.4 km for the readings between 4 and 4.30 a.m., and 0.21 km between 6 and 6.30 p.m. \\

Fig. \ref{fig-8} shows the data on the lifetime $\tau$ of the spatial configurations, defined as $\tau=\xi/<v>$. It is remarkable that the increase of the intensity $A$ of the flow between 6 and 8 a.m. and the minimum of the velocity $<v>$ at 8 a.m. accompanied with sharp maximum of the velocity standard deviation $\sigma$ influence the lifetime $\tau$ only slightly. Another fall of $v$ after 3 p.m. with even sharper peak of $\sigma$ can be correlated with irregularities of the lifetime $\tau$. It is between 6 and 8 p.m. where we observe a qualitative drop of the lifetime $\tau$. Simultaneously, the velocity standard deviation $\sigma$ gets a daily maximum of 25 km/h, as shown in Fig. \ref{fig-5}.

\section{Discussion}

As it follows from our results, the data collected between 6 and 8 p.m. describe states with the lowest mean velocity $<v>$ and the shortest lifetime $\tau$. In the same area, the velocity variance $\sigma$ is the largest. This finding is of interest because of the notion, that the velocity variance shows a maximum at the transition between the flow phase and the congested phase. It was stated in \cite{neu}, that the existence of this maximum is a numerical artifact of an appearance of two different phases within the same time interval. Our interpretation is that the large velocity variance marks a series of relatively rapid changes between different spatial configurations of vehicles. Such a behaviour is perhaps more specific for traffic than an unstable synchronized state. The short lifetime seems to be an apt characteristics of the varying state. It would be desirable to evaluate it also for 
traffic systems without ramps, where the dynamics is not influenced by external sources or sinks.\\

According to the statistics provided by the Insurance Institute for Highway Safety, it is between 5 and 7 p.m. when most people are killed in road accidents \cite{iihs}. As we have seen in the presented data, the road conditions in these hours are: the largest traffic intensity, moderate mean velocity and the largest velocity variance. As noted above, the latter quantity is expected to be low in a synchronized phase of any kind, and - what is obvious - in the congested phase. The fact that between 6 and 8 p.m. the traffic configurations, as seen by drivers, change most quickly, can contribute to the deadly statistics, as these quick changes demand for the drivers' highest attention. It is likely that the lifetime proposed here provides a  criterion of evaluation and identification of potentially dangerous traffic states.\\

Finally the lifetimes of particular traffic states, when handled within theory of diffusion on networks \cite{mejn}, can be converted into rates of variations of time-dependent probabilities of microscopic and coarse-grained states of complex systems. This kind of modelling, proposed recently in \cite{mjk}, seems to us particularly promising in applications to complex systems.

\section*{Acknowledgements} The invaluable help of the Madrid City Council for having provided real data for analysis from the M30 road is gratefully acknowledged. The research is partially supported within the FP7 project SOCIONICAL, No. 231288.

\end{document}